\documentclass[12pt,oneside]{article}
\usepackage{amsfonts,amssymb,graphicx}
\def\be{\begin{equation}}
\def\ee{\end{equation}}
\def\bea{\begin{eqnarray}}
\def\eea{\end{eqnarray}}
\def\c {\bar{c}}

\def\<{\langle}
\def\>{\rangle}
\def\~{\tilde}
\def\s{\sigma}

\def\L{\Lambda}

\def\t{\tau}

\newcommand{\av}[1]{\mbox{{\rm Av}}\left(#1\right)}

\newtheorem{remark}{Remark}
\newtheorem{proposition}{Proposition}

\newtheorem{definition}{Definition}

\newcommand{\beq}{\begin{eqnarray}}
\newcommand{\eeq}{\end{eqnarray}}
\begin{document}
\begin{center}
\vspace{1truecm}
{\bf\Large Stochastic Stability:\\ a Review and Some Perspectives.}\\
\vspace{1cm}
{Pierluigi Contucci}\\
\vspace{.5cm}
{\small Dipartimento di Matematica} \\
{\small Universit\`a di Bologna, 40127 Bologna, Italy}\\
{\small {e-mail: {\em contucci@dm.unibo.it}}}\\
%%%
\vskip 1truecm
\end{center}
\vskip 1truecm
\begin{abstract}\noindent
A review of the stochastic stability property for the Gaussian spin glass models
is presented and some perspectives discussed.
\end{abstract}

\section{Introduction}

In this paper we review the property of {\it stochastic stability} originally introduced
for the Sherrington-Kirkpatrick spin-glass mean field model in \cite{AC}. Here we show 
some of its
consequences expressed in terms of the quenched equilibrium state both in the form
of identities for the overlap distribution and of quenched additivity of the free energy. 
Overlaps between two spin configurations
$\sigma^{(1)}$ and $\sigma^{(2)}$ are usually defined as 
\be\label{overla}
q(\sigma^{(1)},\sigma^{(2)}) \; = \; \frac{1}{N}\sum_{i=1}^{N}\sigma_i^{(1)}\sigma_i^{(2)} \; ,
\ee
but our treatment is given in full generality since it is by now well known \cite{Co} that 
every Gaussian spin glass model has an equilibrium state well expressed by the properties of a probability 
measure of a suitable overlap structure given by its covariance matrix. 
Stochastic stability provided a first and simple 
method to produce an infinite family of identities for the overlap variables. Identities
for random variables with respect to the quenched state reduce the degrees of freedom
of the model and go toward the core of the Parisi mean-field theory: spin glasses are
described by a probability distribution of a single overlap variable and the collections
of copies necessary to describe the whole equilibrium state can be obtained by a suitable
combinatorial rule called {\it ultrametric} which holds for classes of equivalent overlap structures
({\it overlap equivalence}). Although a similar research project is not even completed for the 
Sherrington-Kirkpatrick model important progresses have been done toward it and 
there are clear indications, some based on numerical work \cite{CGGV, CGGPV1,CGGPV2} 
some on rigorous grounds (see the last section), that mean field models and short-range 
finite-dimensional ones behave quite similarly as far as the factorization rules are concerned.
Stochastic stability is deeply rooted also within the physics community. It has
immediately been used in fact in \cite{FMPP1,FMPP2} to determine a relation 
between the off-equilibrium dynamics which is experimentally accessible and the static properties
and is considered, from the theoretical point of view, a structural property
of the spin glass phase \cite{Pa1,Pa2}.

The paper is organized as follows: the first section introduce the basic notions of a spin glass systems 
and the relative notations, the second the property of stochastic stability and its consequences. The third examines
some perspectives in the light of some interesting recent development and provides some perspective.

\section{Definitions}
\label{def}

We consider a disordered model of Ising configurations
$\s_n=\pm 1$, $n\in \Lambda\subset {\cal L}$ for some subset
$\Lambda$ (volume $|\Lambda|$) of some infinite graph ${\cal L}$. We denote
by $\Sigma_\Lambda$ the set of all $\s=\{\s_n\}_{n\in \Lambda}$, and
$|\Sigma_\Lambda|=2^{|\Lambda|}$. In the sequel the
following definitions will be used.

\begin{enumerate}

\item {\it Hamiltonian}.\\ For every $\Lambda\subset {\cal L}$ let
$\{H_\Lambda(\sigma)\}_{\s\in\Sigma_N}$
be a family of
$2^{|\Lambda|}$ {\em translation invariant (in distribution)
Gaussian} random variables defined according to
the  general representation
\be
H_{\Lambda}(\s) \; = \; - \sum_{X\subset \Lambda} J_X\s_X
\label{hami}
\ee
where
\be
\s_X=\prod_{i\, \in X}\s_i \; ,
\ee
($\s_\emptyset=0$) and the $J$'s are independent Gaussian variables with
mean
\be\label{mean_disorder}
{\rm Av}(J_X) = 0 \; ,
\ee
and variance
\be\label{var_disorder}
{\rm Av}(J_X^2) = \Delta^2_X  \; .
\ee
\item {\it Average and Covariance matrix}.\\
The Hamiltonian $H_{\Lambda}(\s)$ has covariance matrix
\begin{eqnarray}\label{cov-matr}
\label{cc}
{\cal C}_\Lambda (\s,\tau) \; &:= &\;
\av{H_\Lambda(\s)H_\Lambda (\tau)}
\nonumber\\
& = & \; \sum_{X\subset\Lambda}\Delta^2_X\s_X\t_X\, .
\end{eqnarray}
The two classical examples are the covariances of the Sherrington-Kirkpatrick model
and the Edwards-Anderson model. A simple computation shows that the first is the square
of the function {\rm (\ref{overla})} and the second is the link-overlap 
\be
\frac{1}{|\Lambda|}\sum_{|i-j|=1}\s_i\s_j\t_i\t_j \; .
\ee
By the Schwarz inequality
\be\label{sw}
|{\cal C}_\Lambda (\s,\t)| \; \le \; \sqrt{{\cal C}_\Lambda
(\s,\s)}\sqrt{{\cal C}_\Lambda (\t,\t)} \; = \;
\sum_{X\subset\Lambda}\Delta^2_X
\ee
for all $\s$ and $\t$.
\item {\it Thermodynamic Stability}.\\
The Hamiltonian (\ref{hami}) is thermodynamically stable if there exists
a constant $\c$ such that
\begin{eqnarray}
\label{thst}
\sup_{\Lambda\subset {\cal L}}
\frac{1}{|\Lambda|}\sum_{X\subset\Lambda}\Delta^2_X
\; & \le & \; \c \; < \; \infty\;.
\end{eqnarray}
Thanks to the relation (\ref{sw}) a thermodynamically stable model fulfills the bound
\begin{eqnarray}
\label{pippo}
{\cal C}_\Lambda (\s,\t) \; & \le & \; \c \, |\Lambda|
\end{eqnarray}
and has an order $1$ normalized covariance
\begin{eqnarray}\label{norm_covar_matrix}
c_{\Lambda}(\s,\t) \; & : = & \; \frac{1}{|\Lambda|}{\cal C}_\Lambda (\s,\t)\;.
\end{eqnarray}
\item {\it Random partition function}.
\be\label{rpf}
{\cal Z}_\Lambda(\beta) \; := \; \sum_{\s  \in \,\Sigma_\Lambda}
e^{-\beta{H}_\Lambda(\s)}
%\equiv \sum_{\s  \in \,\Sigma_\Lambda}e^{-\b\hlp(\s)-\b\hcl(\s)}
\; ,
\ee
\item {\it Random free energy/pressure}.
\be\label{rfe}
-\beta {\cal F}_\Lambda(\beta) \; := \; {\cal P}_\Lambda(\beta) \; := \; \ln {\cal Z}_\Lambda(\beta)
\; ,
\ee
\item {\it Random internal energy}.
\be\label{rie}
{\cal U}_\L(\beta) \; := \; \frac{\sum_{\s  \in \,\Sigma_\Lambda}
H_{\Lambda}(\s)e^{-\beta{H}_\Lambda(\s)}}{\sum_{\s  \in \,\Sigma_\Lambda}
e^{-\beta{H}_\Lambda(\s)}}
\; ,
\ee
\item {\it Quenched free energy/pressure}.
\be
-\beta F_{\Lambda}(\beta) \; := \; P_{\Lambda}(\beta) \; := \; \av{ {\cal P}_{\Lambda}(\beta) }\; .
\ee
%\be
%-\beta F_{\Lambda,\Lambda^\prime}(\beta) \; := \; P_{\Lambda,\Lambda^\prime}(\beta) \; := \; \av{ {\cal P}_{\Lambda,\Lambda^
%\prime}(\beta) }\; .
%\ee
\item {\it Random Boltzmann-Gibbs state}.
\be
\omega(-) \; := \; \sum_{\sigma}(-)
\frac{e^{-\beta H_\Lambda}}{{\cal Z}_\Lambda(\beta)} \; ,
\ee
and its $R$-product version.
\be\label{erre}
\Omega_{\Lambda} (-) \; := \;
\sum_{\sigma^{(1)},...,\sigma^{(R)}}(-)\,
\frac{
e^{-\beta[H_\Lambda(\s^{(1)})+\cdots
+H_\Lambda(\sigma^{(R)})]}}{[{\cal Z}_{\Lambda}(\beta)]^R}
\; .
\label{omega}
\ee
%\item {\it Quenched equilibrium state}.
%\be
%\<-\>_{\Lambda} \, := \av{\Omega_{\Lambda} (-)} \; .
%\ee
\item\label{obs} {\it Quenched overlap observables}.\\
For any smooth bounded function $G(c_{\Lambda})$
(without loss of generality we consider $|G|\le 1$ and no assumption of
permutation invariance on $G$ is made) of the covariance matrix
entries we introduce (with a small abuse of notation) the random
$R\times R$ matrix of elements $\{c_{k,l}\}$ (called {\it generalized overlap}) 
and its measure $\<-\>_\Lambda$ by the formula
\be
\<G(c)\>_\Lambda \; := \; \av{\Omega (G(c_{\Lambda}))} \; .
\ee
E.g.:
$G(c_\Lambda)= c_{\Lambda}(\sigma^{(1)},\sigma^{(2)})c_{\Lambda}(\sigma^{(2)}
,\sigma^{(3)})$
\be
\<c_{1,2}c_{2,3}\>_\Lambda \; = \;
\av{\sum_{\sigma^{(1)},\sigma^{(2)},\sigma^{(3)}}
c_{\Lambda}(\sigma^{(1)},\sigma^{(2)})c_{\Lambda}(\sigma^{(2)},\sigma^{(3)})
\;\frac{
e^{-\beta[\sum_{i=1}^{3}H_\Lambda(\s^{(i)})]}}{[{\cal Z}(\beta)]^3}} \; .
\ee
\end{enumerate}

\section{Stochastic Stability}\label{sec_interp}
Given the Gaussian process $H_\Lambda(\sigma)$ of covariance ${\cal C}_\Lambda(\s,\t)$ we introduce an 
independent Gaussian process, $K_\Lambda(\s)$, defined by the covariance $c_\Lambda(\s,\t)$,  
the deformed random state
\be
\omega^{(\lambda)}_\Lambda (-) \; = \; 
\frac{\omega( - e^{\lambda K_\Lambda})_\Lambda}{\omega(e^{\lambda K_\Lambda})_\Lambda}
\ee
and its relative deformed quenched state $\< - \>^{(\lambda)}_\Lambda = \av{\Omega^{(\lambda)}_\Lambda (-)}$.

\begin{definition}{\bf Stochastic Stability} {\rm \cite{AC,CGi}}\\
A Gaussian spin glass model is stochastically stable if the deformed quenched state 
and the original one do coincide in the thermodynamic limit:
\be
\lim_{\Lambda\nearrow {\cal L}}\< - \>^{(\lambda)}_\Lambda \; = \; \lim_{\Lambda\nearrow {\cal L}} \< - \>_\Lambda
\ee
\end{definition}
\begin{remark}
In spin glass models the existence of the thermodynamic limit has been settled only
at the level of the free energy {\rm \cite{GT, CL}}. For the correlation functions there
are only abstract results using compactness arguments or, equivalently, existence along subsequences. 
In comparison with models without disorder, for instance
the ferromagnetic ones, what is lacking is the control of the local correlation functions
in terms of the interaction parameters. While ferromagnetic correlations increase motononically when
varying any spin interactions (Griffiths, Kelly and Sherman inequalities of type II {\rm \cite{Gr,KS}}) 
nothing is know about a spin glass 
correlation when the interaction distribution is changed. It is known nevertheless that spin glass 
correlation functions are not monotonic in the volume {\rm \cite{CUV}} when the interaction is centered.
For non zero average of the interaction a special result
exists in the Nishimori line where motononicity is recovered {\rm \cite{CMN}}.
\end{remark}
Since the Hamiltonian $H$ and the field $K$ have a mutually rescaled distribution
\be
H_\Lambda(\sigma) \mathrel{\mathop{=}\limits^{\cal D}} \sqrt{|\Lambda|} K_\Lambda(\sigma)
\label{eq:HK}
\ee
the addition law for the Gaussian variables implies
\be
\sqrt{\beta^2 + {\lambda^2\over |\Lambda|}} H(\sigma,J) \
\mathrel{\mathop{=}\limits^{\cal D}} \
\beta H(\sigma,J) + \lambda K(\sigma) \; ,
\label{eq:betalambda}
\ee
i.e. the deformation with a field $K$ is equivalent to a change of 
the order $O(\frac{1}{N})$ in the temperature. The previous identity shows that
the deformed measures do coincide, a part on points of discontinuity with respect
to the temperature, with the original unperturbed one. Let consider some 
consequences of the stochastic stability that have been proved in a series of works.

\begin{proposition}\label{ss} {\bf Zero-average polynomials}.\\
For every monomial $Q$ of the overlap algebra (e.g. $c_{1,2}$, or $c^2_{1,2}c_{2,3}$) 
for a gaussian spin glass model defined by the covariance in (\ref{cov-matr})
the following property hold:
\be
{\partial^2\over \partial \lambda^2} \< Q \>^{(\lambda)}_\Lambda |_{\lambda=0} =
\<\Delta Q \>_\Lambda  \; ,
\label{prop:16}
\ee
where the quantities $\Delta Q$ are polynomials 
which can be computed with a graph-theoretical algorithm or with the standard
Parisi replica limit $n\to 0$ formula (see {\rm \cite{AC}}). Moreover the property 
(\ref{eq:betalambda}) implies
\be
{\partial \over 2\beta\partial\beta} \< Q \>_\Lambda = |\Lambda|
{\partial^2\over \partial \lambda^2} \< Q \>^{(\lambda)}_\Lambda |_{\lambda=0} =
|\Lambda| \<\Delta Q \>_\Lambda \; .
\label{17}
\ee
A simple computation allows to deduce that, for every interval $[\beta_0,\beta_1]$
\be\label{ac}
\lim_{\Lambda\nearrow {\cal L}}\int_{\beta_0^2}^{\beta_1^2}\<\Delta Q \>_\Lambda d\beta^2 \; = \; 0
\ee
i.e. the vanishing in $\beta^2$-average of the quantities 
$\<\Delta Q \>_\Lambda$ when the thermodynamic limit is considered. See also {\rm \cite{Ba}}
for an independent method to obtain the previous result which works for general (including 
non Gaussian) distributions and {\rm \cite{BaGe}} for an interpretation of the identities
as the Noether's conserved quantities in a classical mechanics theory.
\end{proposition}

The consequences seen so far derive from the computation of the first two derivatives
and basically mean that for a stochastic perturbation tuned by a parameter $\lambda$
not only the first derivative vanishes when computed in zero (which is obvious for symmetry
reasons) but also the second one in the thermodynamic limit does i.e. the curvature
of the perturbed state is a vanishing function for increasing volumes. 

The possibility that the computation of higher order derivatives could lead to
new results beyond (\ref{ac}) has been investigated in \cite{Co2,BCK} and has a negative 
answer due to the following result
\be
{\partial^{2n}\over \partial \lambda^{2n}} \< Q \>^{(\lambda)}_\Lambda |_{\lambda=0} = (2n-1)!!
\<\Delta^n Q \>_\Lambda  \; ,
\label{pro:16}
\ee
which implies that the vanishing of higher order derivatives provides the same 
information of the second one at the level of the whole algebra of observables.

In the paper \cite{AC} it has been proved that stochastic stability 
is equivalent to the following property:

\begin{proposition} {\bf Quenched additivity}.\\
Given any finite collection of independent Gaussian fields 
$K_{(1)}(\sigma), K_{(2)}(\sigma),\ldots, K_{(l)}(\sigma)$ (independent 
also on the Hamiltonian)
with the same covariance (\ref{cov-matr}), and any smooth polynomially
bounded functions $F_1, F_2,\ldots, F_l$ a Gaussian spin glass model fulfills
the following relation
\be
\lim_{\Lambda\nearrow {\cal L}} {\rm Av}\; \ln \Omega_\Lambda\left( \exp(\sum_{i=1}^{l} F_i(K_{(i)}))\right) =
\lim_{\Lambda\nearrow {\cal L}} \sum_{i=1}^{l} {\rm Av}\; \ln \Omega_\Lambda\left(\exp F_i(K_{(i)})\right) \ .
\label{eq:log}
\ee
where the Gaussian measure ${\rm Av}(-)$ include the integration on all the fields $K_{(i)}$ and the
Hamiltonian, and the state $\Omega(-)$ is defined in (\ref{erre}).
The previous formula would be trivial if the fields $F_i$ would be independent
with respect to the measure $\Omega$ which is not the case for the class of fields
considered here. Equivalently, denoting the truncated expectations (cumulants) of order $p$ of the Gibbs-Boltzmann
state by $\Omega(-;p)$, the previous relation says that
\be\label{logide}
\lim_{\Lambda\nearrow {\cal L}} {\rm Av}\; \Omega_\Lambda\left(\sum_{i=1}^{l}F_i(K^{(i)});p\right)  \ = \
\lim_{\Lambda\nearrow {\cal L}} \sum_{i=1}^{l} {\rm Av}\; \Omega_\Lambda\left(F_i(K^{(i)});p\right)
\ee
for every integer $p$, i.e. the linearity in average of the truncated correlation functions
of every order.
\end{proposition}

\section{Some perspectives: toward ultrametricity}

The method of stochastic stability is strictly related to the fluctuation method introduced
in \cite{Gu} and developed in \cite{GhiGu}. 
The main idea of that method is to show that from the simple 
control of the fluctuations for the Hamiltonian per particle, which parallels the law of large numbers,
one can deduce a set of identities. The relation between stochastic stability and the
method of the fluctuations is developped in the paper \cite{CGi2} where it was proved that the linear part
of the Ghirlanda-Guerra identities coincide with the identities produced by stochastic stability 
(see also \cite{Ar} for a further interesting result in the framework of the competing particle systems).

The relation between stochastic stability and the method of fluctuation 
is still under investigation, in particular one would like to know if the two
properties suffice to prove a much stronger property called ultrametricity:
an overlap distribution is called ultrametric if it is supported (in the thermodynamic
limit) only on isoceles and equilateral overlap configurations. For such distributions
the measure of scalenes configurations is zero.

Some recent interesting developments are approching the solution of the problem.
To explain them it is important to report a stronger definition of stochastic stability introduced 
in \cite{AC}
\begin{definition}{\it Extended Stochastic Stability}\\
A Gaussian spin glass model is stochastically stable in the extended sense 
if it is stochastically stable under deformation with respect to all $K^{(p)}_\Lambda(\s)$
whose covariance are $c^p_\Lambda(\s,\t)$. Its consequences clearly
extend to every power of the covariance $c$ the results already proved
with the standard stochastic stability.
\end{definition}
In the framework of the competing particle systems it has been proved \cite{ArAi} 
that the extended stochastic stability if applied to overlaps which take only a finte 
number of values implies ultrametricity. 
The same result has been obtained, for the same class of overlap distribution, using 
an extended version of the fluctuation method in \cite{Pa,Ta}.

It is important to stress that the formalism developed so far is not limited to the
mean field spin glasses. Indeed all the results which we have shortly reviewed
hold for a general Gaussian spin glass model in terms of the proper covariance matrix.
In particular the recent developments, in which ultrametricity under certain
hypotheses has been proved, do not distinguish between the mean field case or the short-range
finite-dimensional one like the Edwards-Anderson model. They give indeed a clear indication
that the factorization structure of the two cases is likely to be the same. Of course
even such similarity for the factorization structure wouldn't be enough to guarantee
the same low-temperature phase for the two models. The celebrated Parisi self-consistence 
equation \cite{MPV}, which implies that the overlap distribution of the Sherrington-Kirlpatrick 
model has a non-trivial support, is in fact very specific for the mean field case with its strong 
permutation invariance symmetry. It is still not clear what would be the structure of the 
link-overlap distribution for the Edwards-Anderson model if a complete ultrametric factorization 
would take place.

\vspace{1.cm} \noindent {\bf Acknowledgements:} We thank M.Aizenman, L-P. Arguin, 
A.Barra, A.Bovier, C.Giardina, C.Giberti, S.Graffi, F.Guerra, J.Lebowitz, H.Nishimori, G.Parisi,
S.Starr and C.Vernia for many interesting discussions.

%%%%%%%%%%%%%%%%%%%%%%%%%%%%%%%%%%%%%%%%%%%%%%%%


\begin{thebibliography}{CDGG}

\bibitem[AC]{AC} M.Aizenman, P.Contucci,
``On the Stability of the Quenched state in Mean Field Spin Glass Models'',
J. Stat. Phys., Vol. {\bf 92}, N. 5/6,  765-783, (1998)

\bibitem[Co]{Co} P.Contucci, ``Replica Equivalence in the
Edwards-Anderson Model'',  J. Phys. A: Math. Gen., Vol. {\bf 36},
10961-10966, (2003)

\bibitem[CGGV]{CGGV} P.Contucci, C.Giardina, C.Giberti, C.Vernia 
``Overlap Equivalence in the Edwards-Anderson Model", 
Physical Review Letters, Vol {\bf 96}, 217204 (2006) 

\bibitem[CGGPV1]{CGGPV1} P.Contucci, C.Giardina, C.Giberti, G. Parisi, C.Vernia,  
``Ultrametricity in the Edwards-Anderson Model", 
Physical Review Letters, Vol. {\bf 99}, 057206, (2007)

\bibitem[CGGPV2]{CGGPV2} P.Contucci, C.Giardina, C.Giberti, G.Parisi, C.Vernia 
``On the structure of correlations in the three dimensional spin glasses"
Physical Review Letters, Vol. {\bf 103}, 017201 (2009)

\bibitem[FMPP1]{FMPP1} S.Franz, M.Mezard, G.Parisi, L.Peliti, 
``Measuring equilibrium properties in aging systems"
Physical Review Letters, Vol. {\bf 81}, 1758 (1998) 

\bibitem[FMPP2]{FMPP2} S.Franz, M.Mezard, G.Parisi, L.Peliti, 
``The response of glassy systems to random perturbations: A bridge between equilibrium and off-equilibrium"
Journal of Statical Physics, {\bf 97}, 459 (1999) 

\bibitem[Pa1]{Pa1} G.Parisi
``Stochastic Stability", 
page 73, Disordered and Complex Systems, A.I.P (2001)

\bibitem[Pa2]{Pa2} G.Parisi
``Spin glasses and fragile glasses: Statics, dynamics, and complexity"
Proceedings of the National Academy of Sciences, Vol. {\bf 203}, N. 21, (2006) 

\bibitem[GT]{GT}
F. Guerra, F.L. Toninelli, `` The Thermodynamic Limit in Mean Field
Spin Glass Models'', Commun. Math. Phys. {\bf 230}, 71-79 (2002)

\bibitem[CL]{CL}
P.Contucci, J.Lebowitz, ``Correlation Inequalities for Spin Glasses"
Annales Henri Poincare, Vol. {\bf 8}, N.8, 1461-1467, (2007)

\bibitem[Gr]{Gr} R. B. Griffiths,  
``Correlation in Ising Ferromagnets", 
Journal of Mathematical Physics, Vol. {\bf 8}, 478-483, (1967) 

\bibitem[KS]{KS} D.G.Kelly, S. Sherman 
``General GriffithsÕ Inequalities on Correlations in Ising Ferromagnets", 
Journal of Mathematical Physics {\bf 9}, 466, (1968) 

\bibitem[CUV]{CUV} P.Contucci, F.Unguendoli, C.Vernia, 
``Lack of monotonicity in spin glass correlation functions"
Journal of Physcs A: Math. Theor. Vol. {\bf 41}, (2008)

\bibitem[CMN]{CMN}P.Contucci, S.Morita, H.Nishimori
``Surface Terms on the Nishimori Line of the Gaussian Edwards-Anderson Model"
Journal of Statistical Physics, Vol. {\bf 122}; No. 2, pages 303-312, (2006)

\bibitem[CGi]{CGi} P. Contucci, C. Giardin\`a,
``Spin-Glass Stochastic Stability: a Rigorous Proof''
Annales Henri Poincare Vol. {\bf 6}, No. 5, 915 - 923 (2005)

\bibitem[Ba]{Ba} A.Barra
``Irreducible free energy expansion and overlaps locking in mean field spin glasses"
Journal of Statistical Physics, Vol. {\bf 123}; No. 3, pages 601-614, (2006)

\bibitem[BaGe]{BaGe} A.Barra, G.Genovese
``A mechanical approach to mean field spin models"
Journal of Mathematical Physics, Vol. {\bf 50}, 053303 (2009)

\bibitem[Co2]{Co2} P.Contucci
``Toward a classification theorem for stochastically stable measures"
Markov Processes and Related Fields, Vol. {\bf 9}, N. 2, 167-176, (2002)

\bibitem[BCK]{BCK}A.Bianchi, P.Contucci and A.Knauf
``Stochastically Stable Quenched Measures"
Journal of Statistical Physics, Vol. {\bf 117}, Nos. 5/6, 831-844, (2004)

\bibitem[Gu]{Gu}
F.Guerra, ``About the overlap distribution in a mean field spin glass
model'', Int. J. Phys. B,  Vol. {\bf 10},  1675--1684 (1997)

\bibitem[GhiGu]{GhiGu} S. Ghirlanda, F. Guerra,
``General properties of overlap probability distributions in
disordered spin systems. Towards Parisi ultrametricity'',
J. Phys. A: Math. Gen., Vol. {\bf 31}, 9149-9155 (1998)

\bibitem[CGi2]{CGi2} P. Contucci, C. Giardin\`a,
``The Ghirlanda-Guerra identities''
Journ. Stat. Phys. 126, 917-931 (2007)

\bibitem[Ar]{Ar} L-P. Arguin,
``Competing Particle Systems and the Ghirlanda-Guerra Identities",
Elec. Jou. Prob. , Vol. {\bf 13}, 2101-2117 (2008)

\bibitem[ArAi]{ArAi} L.-P. Arguin, M. Aizenman
``On the structure of quasi-stationary competing particle systems"
Annals of Probability 2009, Vol. {\bf 37}, No. 3, 1080-1113, (2009)

\bibitem[Pa]{Pa} D. Panchenko
``A connection between Ghirlanda-Guerra identities and ultrametricity"
arXiv:0810.0743v3 [math.PR]

\bibitem[Ta]{Ta} M.Talagrand
``Construction of pure states in mean field models for spin glasses"
arXiv:0810.0743v3 [math.PR]

\bibitem[MPV]{MPV} M.M\'ezard, G.Parisi, and M.A.Virasoro, 
``Spin Glass Theory and Beyond". World Scientific, (1987).

\end{thebibliography}
\end{document}